\documentclass[aps,prb,showpacs,a4paper]{revtex4}
\usepackage{graphicx}

\begin{document}





\title{Tunnel junction as a noise probe}
\author{E.B. Sonin}

\affiliation{Racah Institute of Physics, Hebrew University of
Jerusalem, Jerusalem 91904, Israel  \\ and \\
Low Temperature Laboratory, Helsinki University of Technology, FIN-02015 HUT, Finland }

\date{\today}

\begin{abstract}
The paper investigates theoretically effects of noise on low-bias parts of $IV$ curves of tunnel
junctions. The analysis starts from the effect of shot noise from an additional (noise) junction on the
Coulomb blockaded Josephson junction in high-impedance environment. Asymmetry of  shot noise characterized
by its odd moments results in asymmetry of the $IV$ curve of the Josephson junction. At high currents
through the noise junction the $IV$ curve is very sensitive to electron counting statistics. The theory is
generalized on  another type of noise (phase noise of a monochromatic AC input), on a normal Coulomb
blockaded tunnel junction. The effect of shot noise on the $IV$ curve of a superconducting Josephson
junction in low-impedance environment is also analyzed. From this effect one
 can obtain information on the time necessary for an electron to
tunnel through the junction responsible for shot noise.  In summary, the analysis demonstrates, that the 
low-bias part of the $IV$ curves of tunnel junctions can be a sensitive probe of various types of noise. 

\end{abstract}



%



\maketitle

\section{Introduction}

The problem of noise and decoherence plays the important role both for applications and for fundamental
physics. Noise parameters for electron transport provide a valuable information, which is not available from average
current-voltage dependences. This stimulates development of effective methods of noise investigation. In particular, a lot of
attention is devoted to shot noise
\cite{Blanter}, which is related to discreteness of charge transport and yields direct information on carrier charges. 
Efforts of theorists were invested to studying full counting statistics of shot noise, its non-Gaussian
character \cite{Lez} and asymmetry (odd moments)  \cite{Odd,BKN}. They are also objects of intensive
experimental investigations
\cite{Reul,Rez}. 

Recently it was shown that the zero-bias
anomaly of the incoherent Cooper-pairs current in a Coulomb blockaded Josephson junction is very
sensitive to shot noise from an independent source and therefore can be used for noise investigation
\cite{SN}. Further experimental and theoretical investigations \cite{exp,SN-T,Heik} have demonstrated
that this method especially useful for studying asymmetry of shot noise, which is connected with 
its non-Gaussian character and is very difficult for detection by other methods of noise spectroscopy
\cite{Reul,Rez}. In  addition, other methods to use the Josephson junction as a noise probe were discussed.
 Deblock {\em et al.} \cite{Delft} suggested to use the quasiparticle current for noise detection. The
scheme to use the Josephson junction as a threshold detector of electron counting statistics has also been
investigated theoretically and experimentally \cite{TN,Pek,Grab}.  The scheme exploited the effect of
noise on macroscopical quantum tunneling at currents close to the critical current.

The goal of this paper is to extend the theoretical analysis of the effect of noise on the low-bias part of
the $IV$ curve of a mesoscopic tunnel junction. The previous analysis \cite{SN,exp,SN-T} addressed the
effect of shot noise on a Coulomb blockaded Josephson junction in the weak-coupling limit, when 
 the Josephson coupling energy $E_J$ was small compared to the Coulomb energy $E_c=e^2/C$. Here $C$ is the
relevant capacitance. Shot noise originated from a low current through a parallel junction, and its effect
on the $IV$ curve of the Josephson junction had asymmetry connected with the non-Gaussian character of shot
noise. The effect was  insensitive to counting statistics of the noise current since the latter was so low
that the single-electron tunneling events were well separated in time and their correlation was not
essential. On the basis of the fulfilled analysis one could expect  that a Coulomb blockade normal junction
is also able to effectively probe noise, and other types of noise different from shot noise can be
investigated, but in order to check these expectations an additional analysis was needed. The present paper
is to give answers to the following questions:
\begin{itemize}
\item
Can the zero-bias anomaly of the probing junction be sensitive to the counting statistics of electrons tunneling through
the noise source (noise junction) when the current through the noise junction grows?

\item Is the zero-bias anomaly of the Coulomb blockaded  {\em normal} tunnel junction also sensitive to noise similarly to
the Coulomb blockaded  Josephson junction?

\item Is the zero-bias anomaly of the Coulomb blockaded  tunnel junction sensitive to other types of noise different from
shot noise?

\item Can the Josephson junction in the strong coupling limit $E_J\gg E_C$ also probe noise?
\end{itemize}

The paper gives positive answers to all these questions. The content of the paper is the following. Section \ref{sec2}
reminds the previous analysis \cite{SN,exp,SN-T} for the Coulomb blockade Josephson junction in the weak coupling limit $E_J
\ll E_c$ and extends it on the case of higher noise currents, when the zero-bias anomaly is expected to be sensitive to the
electron counting statistics. The analysis confirms this expectation. At high currents the
effect is essentially different for the cases of strictly periodical sequence of pulses and of random pulses governed
by the Poissonian statistics. Section \ref{sec3} considers
another type of noise: phase fluctuations in a monochromatic microwave signal, which result in decoherence of the
signal. A similar phase noise  is well known in quantum optics. It is shown that the effect of the microwave on the
$IV$ curve of the Josephson junction gives direct information on the decoherence time. Section \ref{sec4} addresses the
effect of shot noise on the zero-bias anomaly of a normal tunnel junction. The latter is also very sensitive to shot
noise and phase fluctuations in an AC input, and therefore can be exploited as a noise detector. Section \ref{sec5}  considers the
Josephson junction in the strong coupling limit in low-impedance environment. This case was chosen because it allows to exploit its
duality to the case of the Josephson junction in the weak coupling limit in high-impedance environment, which was considered in the
previous sections. But duality is valid mostly for the equilibrium {Nyquist-Johnson noise. For shot noise
duality does not work, and the effect of noise on the Josephson junction in low impedance environment 
(superconductive regime) is different from that in high impedance environment (Coulomb blockade regime). The
important feature of the low-impedance case is  the perspective to measure the intrinsic tunneling time,
which is discussed in the end of Sec.
\ref{sec5}.

\section{Josephson junction, weak coupling limit, shot noise} \label{sec2}

\subsection{Without shot noise}

 Let us review well known results concerning the
$IV$ curve without shot noise. We assume that the  Josephson coupling
energy $E_J\cos \varphi$ ($\varphi$ is the Cooper-pair phase difference between two banks of the Josephson
junction) is weak in comparison with the Coulomb energy $E_c =e^2/2C$, where
$C$ is the relevant capacitance (see below), and the  Josephson coupling can be considered as a time-dependent
perturbation. Then the
Golden Rule gives for the current  of Cooper pairs \cite{Aver,SZ,IN}:
\begin{equation}
I={\pi eE_J^2 \over \hbar}[P(2eV)-P(-2eV)]~, 
 \label{IP}\end{equation}
where the function
\begin{equation}
P(E)={1\over 2\pi \hbar}\int_0^\infty dt\, \left[e^{iEt/\hbar}\left\langle
e^{i\varphi(t_0)}e^{-i\varphi (t_0-t)}\right\rangle+ e^{-iEt/\hbar}\left\langle
e^{i\varphi (t_0-t)} e^{-i\varphi(t_0)}\right\rangle\right]
  \label{PE} \end{equation}
characterizes the probability to transfer the energy $E>0$ to environment (or to
absorb the energy $|E|$ from environment if $E<0$) at the time $t_0$. In many cases (equilibrium and shot noise included) the
average phase correlators do not depend on the time $t_0$ after averaging, and Eq. (\ref{PE}) is reduced to the Fourier component
of the average phase correlator \cite{IN}:
\begin{equation}
P(E)={1\over 2\pi \hbar}\int_{-\infty}^\infty dt\, e^{iEt/\hbar}\left\langle
e^{i\varphi(t)}e^{-i\varphi (0)}\right\rangle~.
  \end{equation}
However studying the linear response to the AC input in Sec. \ref{AClin} we shall need the more
general expression Eq.~(\ref{PE}).

Since we use the
perturbation theory with respect to $E_J$ we can calculate phase fluctuations neglecting $E_J$, i.e.,
treating the Josephson junction as a capacitor. The crucial assumption in the phase-fluctuation theory (or
$P(E)$ theory) is that the phase fluctuations are Gaussian \cite{IN} and 
\begin{equation}
\langle e^{i\varphi_0(t_0)}e^{-i\varphi_0 (t_0-t)}\rangle \approx e^{J_0(t)} ~,
  \label{Gauss}    \end{equation}
where the phase--phase correlator
\begin{eqnarray}
J_0(t) =\langle[ \varphi_0(t)
-\varphi_0(0)]\varphi_0(0)\rangle=J_R(t)+ iJ_I(t) 
= 2\int_{-\infty}^\infty {d\omega \over \omega}
\frac{\mbox{Re} Z(\omega)}{R_Q}
 \frac{e^{-i\omega t} -1}{1-e^{-\beta \hbar \omega}}
     \label{phaseCor} \end{eqnarray}
is a complex function of time. The subscript 0 points out that the phase fluctuation $\varphi_0$ is determined by the
equilibrium Johnson-Nyquist noise in the environment, i.e., in the electric circuit with the impedance
$Z(\omega)$. Here $R_Q=h/4e^2=\pi \hbar/2e^2$ is the quantum resistance for Cooper pairs and $\beta =1/k_BT$ is the inverse
temperature. Then
\cite{IN}
\begin{equation}
P(E)={1\over \pi \hbar}\mbox{Re} \left\{ \int_0^\infty dt \exp\left[ J_0(t) +
\frac{iEt}{\hbar}\right]\right\}\, , 
    \label{P-E}    \end{equation}
and the current is
\begin{eqnarray}
I=-{2 eE_J^2 \over \hbar^2}\mbox{Im}\left\{\int_0^\infty dt e^{J_0(t)}
 \sin \left(2eVt\over \hbar\right)\right\}~.
 \label{WC}  \end{eqnarray}
The zero-bias conductance is given by
\begin{equation}
G_0=\left.{dI\over dV}\right|_{V\rightarrow 0}=-{4 e^2E_J^2 \over
\hbar^3}\mbox{Im}\left\{\int_0^\infty t\,dt e^{J_0(t)}\right\} \,.
          \label{G-0}\end{equation}

\begin{figure}
  \begin{center}
    \leavevmode
    \includegraphics[width=0.5\linewidth]{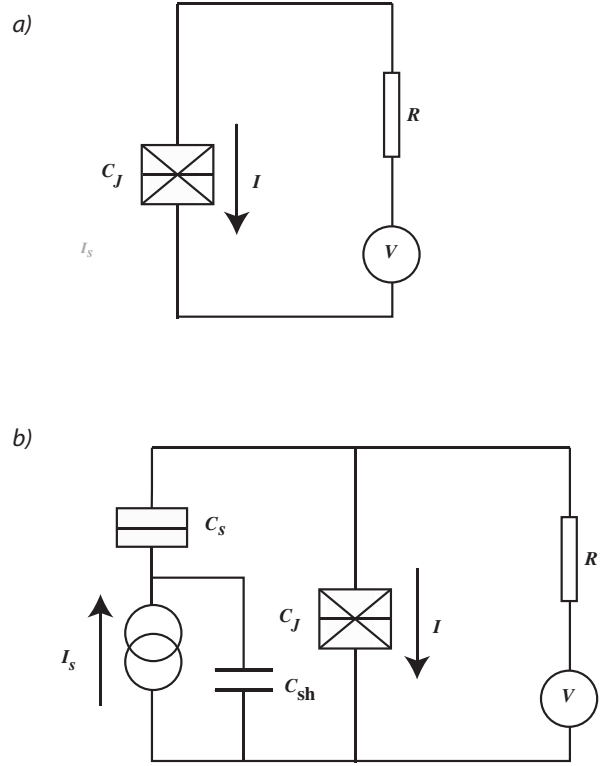}
    \caption{Electric circuit. a) The Josephson junction voltage-biased through the shunt resistance $R$. b) Parallel to
the Josephson junction the normal junction with the capacitance $C_s$ is switched on, which produces shot noise affecting the
current through the Josephson junction. The large shunt capacitance $C_{sh}$ transforms at finite frequencies the current bias
into the voltage bias.}
  \label{fig1}
  \end{center}
  \end{figure}

The electric circuit for our analysis is shown in Fig. \ref{fig1}a. The
Josephson tunnel junction  with capacitance $C_J$ is voltage-biased via the shunt
resistor $R$. Thus in our case  $Z(\omega) =(1/R +i\omega C)^{-1}$ with $C=C_J$, and 
\begin{eqnarray}
J_0(t) = 2\rho\int_{-\infty}^\infty {d\omega \over \omega}
\frac{1}{1 +\omega^2\tau^2}
 \frac{e^{-i\omega t} -1}{1-e^{-\beta \hbar \omega}}~,
     \label{phaseCorT} \end{eqnarray}
where $\tau=RC$ and $\rho=R/R_Q$. At $T \to 0$
($t>0$):
\begin{eqnarray}
J_0(t) =  \rho \left[-e^{t/\tau}\mbox{E}_1\left({t\over \tau}\right)-
e^{-t/\tau}\mbox{E}_1\left(-{t\over \tau}+i0\right) 
-2 \ln{ t \over \tau}
 -2\gamma -i\pi\right] \,,
     \label{J-0}     \end{eqnarray}
 where $\gamma=0.577$ is the Euler constant, and
$\mbox{E}_1(z)=\int_1^\infty (e^{-zt}/t)\,dt$ is the exponential integral
\cite{AS}. The small imaginary correction $+i0$ to the argument of one of the
exponential integrals is important for analytic continuation of $\mbox{E}_1(z)$ from real
$t$ to the complex plane \cite{AS} (see below). At $T=0$ $P(E)$ vanishes for $E<0$ since it is the probability of the transfer of
the energy $|E|$ from the environment to the junction, which is impossible if $T=0$.

For further analysis we need the expressions for $J_0$ in the limits of short time
$t \ll \tau$:
\begin{eqnarray}
J_0(t) \approx  \rho \left[{t^2  \over \tau^2}\left(\ln{t\over
\tau}+\gamma -{3\over 2}\right) -{i\pi t\over \tau}\right]\,,
     \label{J-0s}     \end{eqnarray}
and long time $t \gg \tau$:
\begin{eqnarray}
J_0(t) = - \rho \left(2 \ln{ t \over \tau}
 +2\gamma  +i\pi\right) \, .
     \label{J-0l}     \end{eqnarray}

The function $P(E)$, as well as the current $I$, which it determines, have been carefully studied and calculated for arbitrary
$\rho$ \cite{Ing}. But for the goals of our analysis it is useful to present a simplified calculation valid only in the high
impedance limit  $\rho \gg 1$. In  this limit integration in the expression for the conductance $G_0$, Eq. (\ref{G-0}),
should be done over rapidly oscillating functions and it is difficult to see that the integral exactly  vanishes at $T=0$.
But it becomes evident  after rotation of the integration path in the plane of complex time, which corresponds to replacing
$t$ by $-iy$. This rotation transforms the integration over the positive real semiaxis into the integration over the
negative imaginary semiaxis. After this transformation, the complex function $J_0(t) \to J_0(-iy)$ becomes a
purely real function, and the conductance $G_0$, which is given by the imaginary part of this function, vanishes. This
is true for any term
$V^k$ in the expansion of the current $I$ in voltage $V$ as far as the integral over $t$, which determines this
term, converges at long time. Using the asymptotic expression Eq. (\ref{J-0l}) one can see that the integral is
divergent if
$k>2\rho -2$. So at zero temperature the current $I$ as a function of $V$  does not vanish completely but is given by a
small nonanalytic power-law term, which can be found in the limit of high $\rho$ from the following simple steepest
descent estimation of the integral.

Expecting that the main contribution comes from long times one can rewrite the expression (\ref{P-E}) for the $P(E)$ function
using the asymptotic expression  (\ref{J-0l}) for $J_0$:
\begin{eqnarray}
P(E)= {1\over \pi \hbar}\mbox{Re}\left\{\int _0^\infty dt\,\exp \left[ - \rho \left(2 \ln{ t \over \tau}
 +2\gamma  +i\pi\right) +{iE t\over \hbar}\right]\right\}~.
    \end{eqnarray}
The saddle point is determined by
the condition of vanishing first derivative of the argument of the exponential function: $
-{2\rho / t_0}  + {iE /\hbar}=0$. This yields $t_0=-2i\hbar/E$. Expanding the argument of the exponential
function around the saddle point ($t'=t-t_0$) we receive 
\begin{eqnarray}
 P(E)= {1\over \pi \hbar}\mbox{Re}\left\{\exp\left[\rho\left(-2\gamma -
2\ln {2\rho\hbar\over iE\tau}
-i\pi\right)+2\rho\right]\right\} \int_{-\infty}^\infty
dt'\,\exp\left(-{E^2\over 4\rho\hbar^2} t'^2\right)\nonumber \\
=\exp[2\rho(1-\gamma)]{\tau \over \sqrt{\pi\rho} \hbar} \left(E\tau
\over 2\rho\hbar\right)^{2\rho-1}~.
              \end{eqnarray}
Finally 
\begin{equation}
I={\pi eE_J^2 \over \hbar}P(2eV)=\sqrt{\pi \over {\rho} } \exp[2\rho(1-\gamma)] {eE_J^2 \tau\over \hbar^2}\left(eV\tau \over
\rho\hbar\right)^{2\rho-1}~. 
 \label{curr0}\end{equation}

In summary, if environment provides only the equilibrium noise, the low-bias part of the $IV$ curve is governed by phase
correlations at very long time, which yield in the limit $\rho \gg 1$ an extremely small nonanalytic current.

\subsection{With shot noise}

Now we consider the effect of shot noise from an independent source. Parallel to the Josephson junction there is another junction
(noise junction) with the capacitance $C_s$ connected with an independent
DC current source (see Fig. \ref{fig1}b). The resistance of the noise junction is very large compared to the
shunt resistance $R$. The role of very large capacitance
$C_{sh}$ is to shortcircuit the large ohmic resistance of the current source for finite frequencies. The
current
$I_s$ through the noise junction produces shot noise, which affects the $IV$
curve of the Josephson junction. 

In the presence of shot noise the fluctuating phase $\varphi=\varphi_0+\varphi_s$ consists of
two terms: $\varphi_0$  from Johnson-Nyquist noise, and $\varphi_s$ from shot noise. 
For calculation of the shot-noise fluctuations $\varphi_s$ we assume that
the charge transport through the noise junction is a sequence of current
peaks $\delta I=\mbox{sign}(I_s)e\sum_i\delta (t-t_i)$, where $t_i$ are random
moments of time when an electron crosses the junction \cite{Blanter}. We neglect 
duration of the tunneling event itself. The positive sign of
$I_s$ corresponds to the current shown in Fig. \ref{fig1}b.  Any peak
generates a voltage pulse  at the Josephson junction:
$V_s(t)=\mbox{sign}(I_s)(e/C)\sum_i \Theta (t-t_i) e^{-(t-t_i)/\tau}$, where
$\Theta(t)$ is the step function and $C=C_J+C_s$. The voltage pulses result in phase jumps determined by
the Josephson relation $\hbar \partial \varphi_s/\partial t=2eV_s$. The sequences of current and voltage
peaks and phase jumps are shown in Fig. \ref{fig2}.

Calculating the contribution from shot
noise fluctuations $\varphi_s$ to the phase correlators one should abandon the assumption that noise is Gaussian
[Eq. (\ref{Gauss})]. On the
other hand, we assume that the phase fluctuation
$\varphi_s$ is classical and the values of  $\varphi_s$ at different moments of
time commute. Since equilibrium noise and shot noise are uncorrelated,  the
generalization of Eq. (\ref{WC}) is 
\begin{eqnarray}
I=-{2 eE_J^2 \over \hbar^2}\mbox{Im}\left\{\int_0^\infty dt e^{J_0(t)}
\left\langle \sin \left({2eVt\over \hbar}+\Delta
\varphi_s\right)\right\rangle\right\} .
 \label{curG} \end{eqnarray}

The phase difference between two moments $t_0$ and $t_0-t$, $\Delta \varphi_s=
\varphi_s(t_0)-\varphi_s(t_0-t)=\sum _i 
\delta \varphi_s(t,t_0-t_i)$, is a sum of contributions from random current peaks, each of them determined
by
\begin{eqnarray}
\delta \varphi_s(t,\tilde t) 
=\mbox{sign}(I_s)\pi \rho\left\{ \Theta (\tilde t)\left [ 1-
e^{-(\tilde t)/\tau}\right] 
- \Theta (\tilde t-t)\left [ 1-
e^{-(\tilde t-t)/\tau}\right]\right\}~, 
     \end{eqnarray}
where $\tilde t=t_0-t_i$. Since phase jumps are not  small
for $\rho \gg 1$, one cannot use the perturbation theory with respect to them. 
The interpretation of the expression Eq. (\ref{curG}) for the current is straightforward: As far as the
random phase difference $\Delta \varphi_s=2e\int^{t_0}_{t_0-t}
V_s(t')dt'/\hbar$ is classical,  it should be simply added to the phase difference $\varphi_V=2eVt/\hbar$
generated by the constant voltage bias $V$.

\begin{figure}
  \begin{center}
    \leavevmode
    \includegraphics[width=0.9\linewidth]{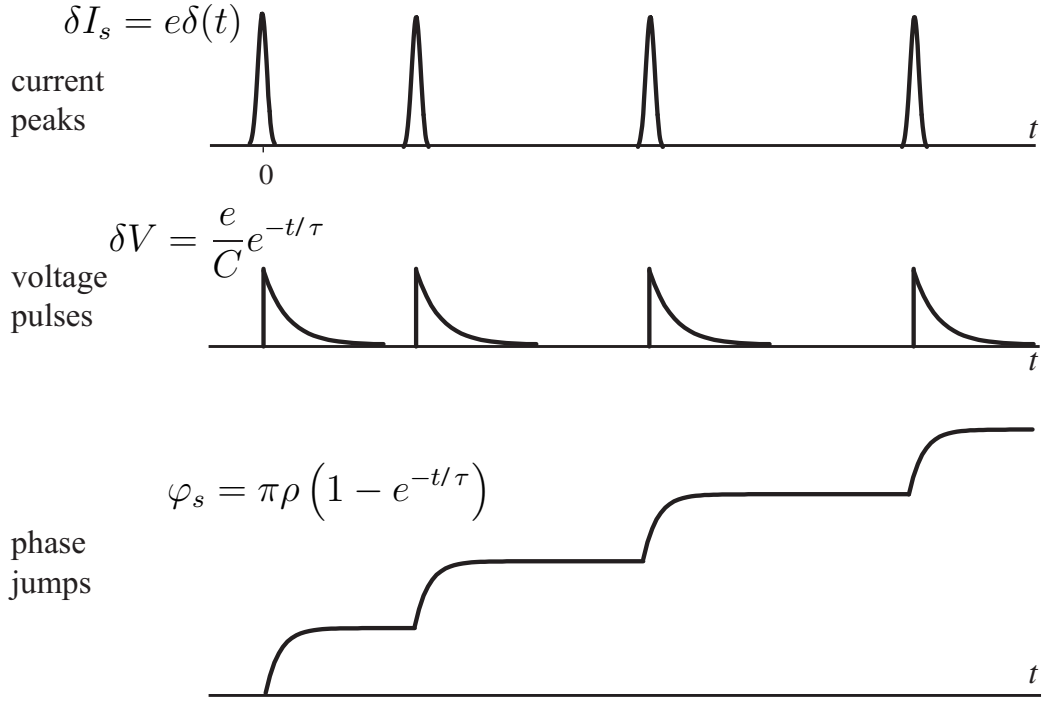}
    \caption{Current and voltage pulses, phase jumps.}
  \label{fig2}
  \end{center}
  \end{figure}

Let us consider the case of the Poissonian statistics. If one performs observation during a very long
period of time
$\tau_\infty$, the number of pulses during this period is large and close to $N\sim |I_s| \tau_\infty/e$. 
Keeping in mind the absence of correlation between pulses, the phase correlator after averaging over
the long time $\tau_\infty$ is
\begin{eqnarray}
 \left\langle e^{i\Delta \varphi_s}\right\rangle =
 \left\langle e^{i\sum _j\delta \varphi_s(t, t_0-t_j) }\right\rangle 
=  \left\langle e^{i\delta \varphi_s }\right\rangle^N 
\approx \left(1+ {\Phi(t)\over \tau_\infty}\right)^N ~,
              \end{eqnarray}
where $\Phi(t)=\Phi_c(t)+i\Phi_s(t)$
is determined by the integrals over a single phase jump (corresponding to a single current pulse):
\begin{eqnarray}
\Phi_s(t)  = \int_{-\infty}^\infty d\tilde
t \sin \delta \varphi_s(t,\tilde t)
= \tau {I_s\over |I_s|} \left\{{\pi \over 2}+ \mbox{si} \left[r\left(1-
e^{-t/\tau} \right) \right]  +\sin r \left[\mbox{ci} r -
\mbox{ci} \left(r  e^{-t/\tau} \right)\right] -\cos r
\left[\mbox{si} r - \mbox{si} \left(r 
e^{-t/\tau} \right)\right] \right\}
        \label{sin}      \end{eqnarray}
and
\begin{eqnarray}
\Phi_c(t)=\int_{-\infty}^\infty d\tilde
t
\left[\cos \delta \varphi_s(t, \tilde t) -1\right] 
= \tau \left\{ \mbox{ci} \left[r\left(1-e^{-t/\tau} \right) \right]+
\cos r
\left[\mbox{ci} r -
\mbox{ci} \left(r  e^{-t/\tau} \right)\right]  \right.\nonumber \\ \left.
+ \sin r
\left[\mbox{si} r - \mbox{si} \left(r 
e^{-t/\tau} \right)\right]     
-\gamma-\ln \left[r\left(1-
e^{-t/\tau} 
\right) \right] \right\}-t~. 
     \label{cos}         \end{eqnarray}
Here $\mbox{si}(x)=-\int_x^\infty
\sin t\, dt/t$ and $\mbox{ci}(x)=-\int_x^\infty \cos t\, dt/t$ are sine and cosine
integral functions \cite{AS}, and $r=\pi \rho$. 
In the limit $\tau_\infty\to \infty$ and $N\to \infty$ at $N/\tau_\infty=|I_s|/e$ the phase correlator is 
\begin{eqnarray}
 \left\langle e^{i\Delta \varphi_s}\right\rangle=  
\exp\left[{N\Phi(t)\over \tau_\infty}\right] =\exp\left[{|I_s|\over e}\Phi(t)\right]~.
      \label{Pois}        \end{eqnarray}
Finally the expression for the current, Eq. (\ref{curG}), can be written as
\begin{eqnarray}
I=-{2 eE_J^2 \over \hbar^2}\mbox{Im}\left\{\int_0^\infty dt e^{J_0(t)}
\exp \left({|I_s| \over e}\Phi_c \right) \sin \left({2eVt\over
\hbar}+{|I_s| \over e} \Phi_s\right) \right\}.
 \label{curGg} \end{eqnarray}

In contrast to the case without shot noise, when the expansion of the current $I$ in a small voltage bias
starts from the nonanalytic term $\propto V^{2\rho-1}$, in the presence of shot noise
the expansion in  $V$ starts with analytic terms:
\begin{equation}
I=I_0 +G_s V+ a V^2+bV^3~,
    \label{Vexp}  \end{equation}
where 
\begin{eqnarray}
G_s=-{ 4e^2 E_J^2 \over \hbar^3 }\int_0^\infty t\, dt \mbox{Im}  \left\{e^{J_0(t)}\right\}[\left\langle\cos
\Delta\varphi_s\right\rangle-1]
               \end{eqnarray}
is the shot-noise conductance, 
\begin{eqnarray}
I_0=-{ 2e E_J^2 \over \hbar^2 } \int_0^\infty  dt \mbox{Im}  \left\{e^{J_0(t)}\right\}
\left\langle\sin \Delta\varphi_s \right\rangle
      \label{ratchet}         \end{eqnarray}
is the ratchet current, and the parameters 
\begin{eqnarray}
a={ 4e^3 E_J^2 \over \hbar^4}\int_0^\infty t^2\, dt \mbox{Im}  \left\{e^{J_0(t)}\right\}
\left\langle\sin \Delta\varphi_s \right\rangle
               \end{eqnarray}
and
\begin{eqnarray}
b={ 8 e^4 E_J^2 \over 3 \hbar^5} \int_0^\infty t^3\, dt \mbox{Im}  \left\{e^{J_0(t)}\right\}[\left\langle\cos
\Delta\varphi_s\right\rangle-1]
               \end{eqnarray}
determine the curvature of the  conductance-voltage plot and the shift of the conductance minimum. Without
shot noise all these integrals vanish at $\rho \gg 1$ [see the paragraph after Eq. (\ref{J-0l})].

In Refs. \onlinecite{exp,SN-T} the parameters of the analytic expansion of $I(V)$ were calculated for low
currents $|I_s|\ll e/\tau$ through the noise junction, when voltage pulses and phase jumps are well
separated in time (see Fig.
\ref{fig2}). Then  the relevant phase correlators are proportional to the density
$|I_s|/e$ of pulses in time \cite{SN-T}:
\begin{equation}
\langle e^{i \Delta \varphi_s} \rangle -1= \langle \cos \Delta \varphi_s \rangle-1
+i\langle \sin \Delta \varphi_s \rangle = {|I_s| \over e}\Phi(t)={|I_s| \over e}[\Phi_c(t)+i\Phi_s(t)]~.
\end{equation}
 The parameters 
determined by $\left\langle\cos \Delta\varphi_s\right\rangle$ correspond to ``even'' effects (the conductance
is symmetric with respect to voltage inversion $V\to -V$), which are present also without shot noise. Therefore the
contribution from shot noise to these parameters  should be
added to the values derived from Gaussian equilibrium noise. In contrast, the parameters determined by
$\left\langle\sin \Delta\varphi_s \right\rangle$  correspond to ``odd'' (asymmetric) effects, which are absent for 
equilibrium noise and are related to the non-Gaussian character of shot noise.

In the high-impedance limit $\rho \gg 1$ it is possible to calculate the parameters
of the $IV$ curve analytically. As one can see below, the most important
contributions to the integrals, which determine $G_s$, $a$, and $b$, come from times $t  \sim \tau
/\sqrt{\rho \ln \rho}$ short compared to $\tau $, and one may use the small-argument expansion for the
Johnson-Nyquist correlator given in Eq. (\ref{J-0s}). On the other hand, these times are long enough for using
asymptotic expansions for the sine and the cosine integrals:
$\mbox{si}(x) \sim -\cos x/x -\sin x/x^2 $, $\mbox{ci}(x) \sim \sin x/x -\cos x/x^2$. Then one can rewrite Eq. (\ref{sin})  as 
\begin{eqnarray}
\Phi_s(t) \approx \tau \left\{{\pi \over 2}+ {1\over r} -{\cos \left[r\left(1-
e^{-t/\tau} \right) \right] \over re^{-t/\tau}}-{\cos \left[r\left(1-
e^{-t/\tau} \right) \right] \over r(1-e^{-t/\tau})} \right.  \nonumber \\ \left.
+{\sin \left[r\left(1-
e^{-t/\tau} \right) \right] \over r^2e^{-2t/\tau}}-{\sin \left[r\left(1-
e^{-t/\tau} \right) \right] \over r^2(1-e^{-t/\tau})^2} \right\}
\approx \tau \left[{\pi \over 2}-{\tau\over rt} \cos {rt\over \tau}  -{\tau^2\over r^2t^2} \sin {rt\over
\tau}
\right]  ~.
      \label{sinAs}      \end{eqnarray}
In Eq. (\ref{cos}) it is enough to keep only the main asymptotic terms $\propto 1/r$:
\begin{eqnarray}
\Phi_c(t) \approx \tau \left\{-{t \over\tau} 
+{\sin \left[r\left(1-e^{-t/\tau} \right) \right] \over re^{-t/\tau}}+{\sin \left[r\left(1-e^{-t/\tau} \right) \right] \over
r(1-e^{-t/\tau})}
 -\gamma-\ln \left[r\left(1-e^{-t/\tau} \right) \right] \right\}\nonumber \\
\approx \tau \left[-{t \over\tau} -\ln {rt\over \tau}  -\gamma
+{\tau \over rt}\sin {rt\over \tau} \right]~. 
          \label{cosAs}  \end{eqnarray}

Let us consider first the parameters connected with ``even'' effects. For low currents $|I_s| \ll
e/\tau $:
\begin{eqnarray}
G_s\approx -{ 4e E_J^2 \tau \over \hbar^3 }|I_s|\int_0^\infty t\, dt \mbox{Im}
\left\{e^{J_0(t)}\right\}{\tau \over rt}\sin {rt\over \tau}  \nonumber \\ 
\approx { 4e E_J^2 \tau^2\over \hbar^3 r}|I_s| \int_0^\infty  dt
\exp \left(-\rho{t^2  \over \tau^2}\ln{\tau \over
t}\right) \sin ^2 {rt \over\tau} 
\approx  {\sqrt{2} \pi e E_J^2 \tau^3\over \hbar^3 r^{3/2}\sqrt{\ln r}}|I_s|
\approx {\pi^{5/2}E_J^2 C^3 \over 4
e^5 \sqrt{2\ln \rho}}\rho^{3/2}|I_s|~,
    \label{integr}           \end{eqnarray}
\begin{eqnarray}
b={ 8 e^3 E_J^2\tau \over 3 \hbar^5}|I_s| \int_0^\infty t^3\, dt \mbox{Im}  \left\{e^{J_0(t)}\right\}{\tau \over rt}\sin {rt\over
\tau} 
\nonumber \\
\approx -{ 8 e^3 E_J^2\tau^2 \over 3 \hbar^5 r}|I_s| \int_0^\infty t^2\, dt \exp \left(-\rho{t^2  \over \tau^2}\ln{\tau \over
t}\right) \sin ^2 {rt \over\tau} \approx -{ 2\pi^2\sqrt{2} e^3 E_J^2\tau^5 \over 3 \hbar^5 r^{5/2} (\ln r)^{3/2}}|I_s|
\approx -{\pi^2 C^2\rho\over 6 e^2\ln \rho}G_s ~.
       \label{integr-b}        \end{eqnarray}
The main contributions to these integrals come from the last term in the asymptotic expansion Eq.
(\ref{cosAs}). The  other terms either vanish (term $\propto \gamma$) or are of higher orders (terms $\propto
t$ and $\propto \ln t$) with respect to $1/r$. A negative sign of $b$ means that in the center of the
zero-bias anomaly the curve ``conductance vs. voltage'' has a maximum but not a minimum. However this is
possible to see only at very low temperatures since finite temperatures give a positive contribution to the
curvature parameter $b$.

In a similar way one can calculate the asymmetry integral 
\begin{eqnarray}
a=-{ 4e^2 E_J^2 \tau \over \hbar^4}I_s\int_0^\infty t^2\, dt \mbox{Im}  \left\{e^{J_0(t)}\right\}
{\tau^2\over r^2t^2} \sin {rt\over \tau}={I_s \over |I_s|}{  C\over 2e }G_s ~.
    \label{integr-a}      \end{eqnarray}
Here the main contribution comes from the last term of the asymptotic expansion Eq.  (\ref{sinAs}). 

 For the integral in Eq.
(\ref{ratchet}), which determines the ratchet current $I_0$, the relevant times are shorter than for the
other integrals:
$t
\sim \tau/\rho \sim R_QC$. Therefore one cannot use the asymptotic expansion Eq. (\ref{sinAs}). Instead one can
integrate by parts, and for low noise currents: 
\begin{eqnarray}
I_0=-{ 2e E_J^2 \over \hbar^2 } {|I_s|\over e} \int_0^\infty  dt \mbox{Im}  \left\{e^{J_0(t)}\right\}
\Phi_s(t) \approx { 2 E_J^2 \over \hbar^2 }|I_s| \int_0^\infty  dt \Phi_s(t) \sin {r t\over \tau}
 = { 2 E_J^2\tau \over \hbar^2 r } |I_s| \int_0^\infty  dt \frac {d\Phi_s(t)}{dt}\cos {r t\over \tau}
\nonumber \\
\approx  { 2E_J^2\tau^2 \over \hbar^2 r }I_s \int_0^\infty  {dt\over t} \cos {r t\over \tau}
\sin {r t\over \tau}={ \pi E_J^2\tau^2 \over 2 \hbar^2 r }I_s= {\pi^2 E_J^2 C^2\over 8e^4 }\rho I_s\, .
     \label{integr-r}     \end{eqnarray}

\subsection{Shot noise from high currents and electron counting statistics} \label{sec2c}

Shot noise is non-Gaussian and asymmetric, but in the case of low noise currents its effect does not
depend on statistics of electron transport and is determined only by density of current pulses in time.
Even if pulses were not random and formed a strictly periodical sequence the effect would be the same. In
order to obtain information on counting statistics from measurements of the $IV$ curve the noise current
should be not small compared with the width of the voltage pulse $
\sim \tau$. One could expect \cite{SN-T} that in the limit of high noise current $I_s$ discreteness of the
electron transport through the junction would be less and less essential and eventually the effect of the
current
$I_s$ would be reduced to the constant voltage drop $V_s =I_s  R$ on the shunt resistor $R$ resulting
in a trivial shift of the voltage applied to the Josephson junction. If it were the case, the long-time asymptote for
the phase correlator would be
\begin{eqnarray}
 \langle e^{i\Delta \varphi_s }\rangle \to \exp \left(2ie V_s t\over \hbar \right)
=\exp \left(2ie I_s Rt\over \hbar \right)=\exp \left(ir {I_s t\over e} \right)~.
          \label{asyTr}    \end{eqnarray} 
This asymptotic behavior really takes place for a strictly periodic sequence of current pulses with period
$T_0=e/|I_s|$. In this case the phase variation for the time interval $nT_0 <t
< (n+1)T_0$ ($n$ is integer) is:
\begin{eqnarray}
\varphi(t)=r \left[n +{1-e^{-(t-nT_0)/\tau} \over1-e^{-T_0/\tau}}  \right]  ~.           
\end{eqnarray}
In the limit $|I_s| \to \infty$ ($T_0 \to 0$) the maximum deviation of the phase from the linear function
$\varphi(t)=rt /T_0$ is of the order $r T_0^2/\tau ^2$, i.e., vanishes in this limit. This yields Eq. (\ref{asyTr}).

However, the expression (\ref{Pois}) derived from the Poissonian statistics has another asymptotic
behavior for $t \to \infty$:
\begin{eqnarray}
 \langle e^{i\Delta \varphi_s }\rangle-1
\to  \exp\left[ \left(e^{ir}-1\right){I_s t\over e}\right] ~.
        \label{asyPoi}       \end{eqnarray}
Thus the $IV$ curve of the
Josephson junction essentially depends on whether the current pulses through the parallel (noise) junction are strictly
periodical in time, or completely random.  This proves strong sensitivity  to counting statistics.

Note that the right-hand side of Eq. (\ref{asyPoi}) coincides with the characteristic function
$\chi(r)=\sum_n P_ne^{ir n}$ for the Poissonian distribution $P_n=e^{-\bar{n}}\sum_n \bar{n}^n/n!$ with
$\bar{n}=I_st/e$. This is a natural result since  the number of phase jumps $n=\Delta \varphi_s/2\pi$ during a
long time $t \gg \tau$ coincides with the number of electrons transmitted through the noise junction during
the same time interval. Equation (\ref{asyPoi})  leads to a conclusion that in the case of the Poissonian
statistics shot noise affects the $IV$ curve even at high noise currents when one might expect suppression of
noise effects. This suppression takes place for usual methods of noise detection probing voltage fluctuations.
Indeed calculating the voltage variance $\langle\Delta V^2 \rangle$ for the Poissonian statistics,
\begin{eqnarray}
 \langle \Delta V^2 \rangle=  \langle  V^2 \rangle- \langle  V \rangle^2
= {e\over 2 C}\langle  V \rangle~,
              \end{eqnarray} 
 one obtains that the relative mean-square-root voltage fluctuation $ \sqrt{\langle \Delta V^2 \rangle
}/\langle  V \rangle$ vanishes at $V_s =\langle  V \rangle=I_s R \to \infty$. However, the
Coulomb blockaded junction probes not voltage but phase, which is much more sensitive to
fluctuations. This ``supersensitivity'' of the Coulomb blockaded junction to fluctuations explains why the
effects of higher moments (or cummulants) are much stronger than in usual methods of noise detection
\cite{BKN,Reul,Rez}.

\section{Josephson junction, weak coupling limit, phase noise in the AC input} \label{sec3}

Now we want to demonstrate that the Coulomb blockaded junction can effectively probe not only shot noise
but other types of noise as well. As an example, phase fluctuations in a monochromatic
electromagnetic wave will be considered. This type of noise is very important for lasers \cite{L}, but the
present analysis addresses  microwaves. Suppose that some microwave source provides an input signal described by phase variation at
the Josephson junction:
\begin{eqnarray}
 \varphi_i=\varphi_0 \sin \psi(t)=\varphi_0 \sin \left[\omega t+ \sum_i \gamma_i
\Theta(t-t_i)\right]~,
        \label{phN}      \end{eqnarray}
where $\gamma_i$ are random phase jumps at random moments of time $t_i$. Frequency of phase jumps is characterized by the
average time $\tau_c$ between the jumps. In fact $\tau_c$ is nothing else as the decoherence time of the microwave signal. We
shall see that the $IV$ curve of the Josephson junction is very sensitive to the decoherence process. But in order to demonstrate
it one should first consider the response of the Josephson junction to  a strictly monochromatic noiseless microwave
input.

\subsection{Response to a monochromatic AC input} \label{AClin}

For a strictly monochromatic signal the phase  difference between the moments $t_0$ and $t_0-t$ is 
\begin{eqnarray}
\Delta \varphi_i(t_0,t)=\varphi_i(t_0)-\varphi_i(t_0-t)=\varphi_0 [\sin \omega t_0-\sin \omega (t_0-t)]
=\varphi_0 [\sin \omega t_0(1-\cos \omega t)+\cos \omega t_0\sin \omega t]~.
              \end{eqnarray}
In full analogy with the previous cases, the current is given by
\begin{eqnarray}
I=-{2 eE_J^2 \over \hbar^2}\mbox{Im}\left\{\int_0^\infty dt e^{J_0(t)}
\sin \Delta
\varphi_i(t_0,t)\right\}~, 
       \end{eqnarray}
but now averaging over the moment $t_0$ is not done. The linearization with respect to the input gives
\begin{eqnarray}
I\approx -{2 eE_J^2 \over \hbar^2}\mbox{Im}\left\{\int_0^\infty dt e^{J_0(t)}
\Delta
\varphi_i(t_0,t)\right\} 
= -\varphi_0{2 eE_J^2 \over \hbar^2}\mbox{Im}\left\{\int_0^\infty dt e^{J_0(t)}
[\cos \omega t_0\sin \omega t +\sin \omega t_0(1-\cos \omega t)]\right\} 
 \label{curAC} \nonumber \\
= I_1 \cos \omega t_0 +I_2 \sin \omega t_0 ~.
   \label{curPh}   \end{eqnarray}
Thus the tunneling current of Cooper pairs consists of two parts: one is out of phase and another is in phase with the
input voltage $ V_i=V_0 \cos \omega t$ (the reactive and the active responses)  where $V_0=\hbar \omega \varphi_0/2e $. Comparing
the active response,
\begin{eqnarray}
I_1= -\varphi_0{2 eE_J^2 \over \hbar^2}\mbox{Im}\left\{\int_0^\infty dt e^{J_0(t)}
\sin \omega t\right\} ~,
              \end{eqnarray}
with Eq. (\ref{WC}) one can see that the ohmic current as a function of frequency can be directly received from the 
$IV$ curve $I(V)$  replacing voltage $V$ by $\hbar \omega/2e$. Therefore the  active (ohmic)
response is given by
\begin{eqnarray}
I_1= \varphi_0 I\left(\hbar \omega \over 2e\right)
= V_0 {2e  \over \hbar \omega} I\left(\hbar \omega \over 2e\right)~.
              \end{eqnarray}
Thus the ohmic AC linear conductance of the junction is a nonanalytic function of frequency proportional to $
\omega^{2\rho -2}$. 

The reactive response for frequencies $\omega \ll \rho/\tau=1/R_QC$ is 
\begin{eqnarray}
I_2=-\varphi_0{2 eE_J^2 \over \hbar^2}\mbox{Im}\left\{\int_0^\infty dt e^{J_0(t)}
(1-\cos \omega t)\right\}\approx -\varphi_0{ eE_J^2 \over
\hbar^2}\mbox{Im}\left\{\int_0^\infty dt e^{J_0(t)} \omega^2 t^2\right\}~ .
      \end{eqnarray}
After rotation in the complex plane ($t \to-iy$):
\begin{eqnarray}
I_2=-\varphi_0{eE_J^2 \over \hbar^2}\omega^2\int_0^\infty y^2 dy e^{-\pi \rho y/\tau}
=-\varphi_0{2 eE_J^2 \over \hbar^2}\omega^2\left(\tau\over \pi \rho\right)^3
=-V_0{2e\over \hbar \omega}{2 eE_J^2 \over \hbar^2}\omega^2\left(R_Q C\over \pi
\right)^3=-V_0\omega C{E_J^2 C^2\over 2e^4}~.
         \end{eqnarray}
The linear dependence on frequency points out that this current corresponds to the capacitance element of the
electrical circuit. Indeed, the effective circuit for the Josephson junction has the capacitance $C$ in parallel to the
tunneling conductance element. But in the presence of the Josephson coupling the geometric capacitance should be
replaced by the effective capacitance (like the mass of an electron in a periodic potential should be replaced by the
effective mass, see, e.g., Ref. \onlinecite{SZ}). The current component $I_2$ is related with the difference between the
effective and the geometric capacitance in the weak coupling limit. In order to check it one should calculate the
second-order (with respect to the Josephson coupling) correction to the Coulomb energy $Q^2/2C$:
\begin{eqnarray}
E={Q^2\over 2 C}+ \sum _{Q_G\neq 0}{V(Q_G)^2 \over E_0(Q)-E_0(Q+Q_G)}~,
         \end{eqnarray}
where summation is over the inverse-lattice ``wave-numbers'' (charges) $Q_G=2e n$ ($n$ is any integer).  For weak
Josephson coupling only two terms  $ Q_G=\pm 2e$ of the sum are important, and $V(\pm 2e)=E_J/2$. Then 
\begin{eqnarray}
E={Q^2\over 2 C}+ {E_J^2\over 4}\left[{2 C \over Q^2-(Q+2e)^2}+ {2C \over
Q^2-(Q-2e)^2}\right]
={Q^2\over 2 C}+{E_J^2\over 4}{e^2 C\over e^4 -e^2Q^2}\approx {Q^2\over 2 C}+{E_J^2 C\over 4e^2}-{E_J^2 C\over
4e^4}Q^2~.
         \end{eqnarray}
This yields for the effective capacitance:
\begin{eqnarray}
{1\over C^*}= {d^2 E\over dQ^2}={1\over C}-{E_J^2 C\over 2e^4}~,
        \end{eqnarray}
and it is evident that the reactive component $I_2$ of the tunneling current is determined by $\Delta C=C^*-C$. It is
useful to rewrite the expression for this correction using  the
Ambegoakar-Baratoff relation:
\begin{eqnarray}
 E_J={\pi \hbar \Delta \over 4e^2 R_T}={\Delta_0 \over 2}{R_Q\over R_T}~,
        \end{eqnarray}
where $\Delta_0$ is the bulk superconducting gap. Then the capacitance correction is
\begin{eqnarray}
\Delta C=C {E_J^2 C^2 \over 2e^4}=C{R_Q^2\over R_T^2}{\Delta_0^2 \over 8e^4/C^2}~.
  \label{Ccor}       \end{eqnarray}

Now we calculate the nonlinear (quadratic) effect of the AC input on the DC conductance. This requires averaging (over
the period of the AC input) of the quadratic term in the expression for the conductance:
\begin{eqnarray}
\Delta G=-{4 e^2E_J^2 \over \hbar^3}\mbox{Im}\left\{\int_0^\infty t\,dt e^{J_0(t)}
\langle \cos \Delta\varphi_i \rangle\right\} 
\approx {2 e^2E_J^2 \over \hbar^3}\mbox{Im}\left\{\int_0^\infty t\,dt
e^{J_0(t)}\langle  \Delta \varphi_i^2 \rangle\right\} \nonumber \\
=\varphi_0^2{2 e^2E_J^2 \over \hbar^3}\mbox{Im}\left\{\int_0^\infty t\,dt
e^{J_0(t)}(1-\cos \omega t)\right\}~.
   \label{DG}   \end{eqnarray}
The expansion in $\omega$ contains only the terms even in $t$. The integrals over these terms vanish after rotation in
the plane of complex time. This means that $\Delta G$ is determined by a very small nonanalytic contribution from long
times \cite{FBS}. The presence of
noise essentially changes the situation as we shall see in the next subsection.

\subsection{Phase noise}

In the presence of the phase noise, when the phase varies according to Eq. (\ref{phN}), the squared difference between the phases
at the moments of time $t_0$ and
$t_0-t$ averaged over the time $t_0$ is given by
\begin{eqnarray}
\langle \Delta \varphi_i(t_0,t)^2\rangle =\varphi_0^2 \langle[\sin \psi(t_0)-\sin \psi
(t_0-t)]\rangle^2 =\varphi_0^2 \{1-\langle\cos[\psi(t_0)-\psi (t_0-t)]\rangle\}\, .
    \end{eqnarray}
The average $\langle\cos[\psi(t_0)-\psi (t_0-t)]\rangle$ does not vanish only if there is no random phase jumps in the time
interval $t$. Assuming the  Poisson distribution the probability of the absence of random phase jumps in the interval $t$
is
$e^{-t/\tau_c}$, where
$\tau_c$ is the average period between random phase jumps. Then one obtains that
\begin{eqnarray}
\langle \Delta \varphi_i(t_0,t)^2\rangle  =\varphi_0^2(1- \cos \omega t e^{-t/\tau_c})~.
    \end{eqnarray}
Finally the correction to the DC conductance [compare with Eq. (\ref{DG})] is:
\begin{eqnarray}
\Delta G=\varphi_0^2{2 e^2E_J^2 \over \hbar^3}\mbox{Im}\left\{\int_0^\infty t\,dt
e^{J_0(t)}(1-\cos \omega t e^{-t/\tau_c})\right\}
\approx \varphi_0^2{2 e^2E_J^2 \over \hbar^3 \tau_c}\mbox{Im}\left\{\int_0^\infty
t^2\,dt e^{J_0(t)}\right\}~.
       \end{eqnarray}
After rotation in the complex plane the imaginary part of the integral does not vanish, and
\begin{eqnarray}
\Delta G=\varphi_0^2{2 e^2E_J^2 \over \hbar^3 \tau_c}\int_0^\infty
y^2\,dy e^{-\pi \rho y/\tau}=\varphi_0^2{4 e^2E_J^2 \tau^3\over\pi^3
\rho^3 \hbar^3 \tau_c}=\varphi_0^2{E_J^2 C^3\over 2 e^4 \tau_c}
=V_0^2{2E_J^2 C^3\over  e^2 \hbar ^2 \omega^2 \tau_c}~.
      \end{eqnarray}
This contribution to the zero-bias conductance essentially exceeds the small nonanalytic contribution from the
strictly monochromatic microwave and provides direct information on the decoherence time $\tau_c$. 

\section{Normal junction, shot noise}  \label{sec4}

\subsection{The $IV$ curve in the time-domain formulation (without shot noise)}

This subsection reminds the $P(E)$ theory for a normal Coulomb blockaded junction \cite{IN}. The current is given
by 
\begin{equation}
  I= { e}[\Gamma^+(V)-\Gamma^-(V)] ~,
        \end{equation}
where the forward tunneling probability is
\begin{equation}
\Gamma^+(V) = {1\over e^2 R_T}\int_{-\infty}^\infty dE \frac{E}{1
-\exp\left(-\frac{E}{T}\right)}P(eV-E)~.
    \label{ProbE} \end{equation}
Here $P(E)$ is given by Eq. (\ref{P-E}). The backward tunneling rate is $\Gamma^- (V)=\Gamma^+(-V)$.
Since now we consider tunneling of single electrons (instead of Cooper pairs) the correlation function is given by
\begin{equation}
J_0(t) =\langle[ \varphi(t) -\varphi(0)]\varphi(0)\rangle
= 2\int_{-\infty}^\infty {d\omega \over \omega} \frac{\mbox{Re} 
Z(\omega)}{R_K} \frac{e^{-i\omega t} -1}{1-e^{-\beta \hbar \omega}}~,
      \end{equation}
where the single electron quantum resistance $R_K=h/e^2$ replaces the Cooper-pair quantum resistance $R_Q=h/2e^2$ in
Eq.~(\ref{phaseCor}). In the case of purely ohmic environment, $Z(\omega)=R/(1+i\tau \omega)$, 
one can use the time-domain formulation \cite{JED} fulfilling integration over energy $E$ in Eq. (\ref{ProbE}):
\begin{equation}
\Gamma^+(V) = {1\over \pi \hbar e^2 R_T}\mbox{Re}\left\{\int_0^\infty dt
\gamma(t)\exp\left[ J_0(t) + \frac{ieVt}{\hbar}\right] \right\}  ~.
    \label{ProbT}\end{equation}
Here $\tau=RC$ as before, and
\begin{equation}
\gamma(t) = \int_{-\infty}^\infty dE 
\frac{E}{1-\exp\left(-\frac{E}{T}\right)}
\exp\left( - \frac{iEt}{\hbar}\right) =i\pi\hbar ^2 {d\over dt} \delta(t)
- {\pi^2 \over \beta^2} \frac{1}{\sinh ^2 (\pi t/ \hbar \beta)}~.
   \end{equation}
Then after some simple algebra the current is 
\begin{eqnarray}
  I= { e}[\Gamma^+(V)-\Gamma^-(V)]= {1\over R_T} \left[
V  + {2\pi \over e\hbar \beta^2} \mbox{Im}\left\{  \int_0^\infty 
\frac{dt}{\sinh ^2 (\pi t/ \hbar \beta)} e^{J_0(t)}\sin \frac{eVt}{\hbar}\right\} \right] ~.
        \end{eqnarray}
At $T=0$:
\begin{eqnarray}
  I= {1\over R_T} \left[V  + {2 \hbar \over \pi e } \mbox{Im}\left\{ \int_0^\infty 
\frac{dt}{t ^2 } e^{J_0(t)}\sin \frac{eVt}{\hbar} \right\}\right] ~,
    \label{i-v}    \end{eqnarray}
and the differential conductance is
\begin{equation}
 G= {dI \over dV}=  {1\over R_T} \left[
1 + {2\over \pi } \mbox{Im}\left\{ \int_0^\infty 
\frac{dt}{t} e^{J_0(t)}\cos \frac{eVt}{\hbar}\right\} \right] ~.
        \end{equation}
For $\rho=R/R_K \gg 1$ in the limit $V \to 0$ the conductance vanishes:
\begin{equation}
 G_0= {1\over R_T} \left[1  - {2\over \pi }  \int_0^\infty \frac{dt}{ t}\sin \left( \pi \rho {t\over
\tau}\right)
 \right] \to  0~,
          \end{equation}
but there remains a small nonanalytic contribution, which originates from long times: $G_0 \propto V^{2\rho}$.

\subsection{Effect of shot noise}

Repeating derivation of the  effect of the random phase from shot noise, which was done above for the Josephson junction,
one receives that the effect on the normal tunnel junction is similar: the  linear phase from the constant voltage,
$eVt/\hbar$, must be replaced  by the phase $eVt/\hbar + \Delta \varphi_s$, where $\Delta \varphi_s=\varphi_s(t_0)
-\varphi_s(t_0-t)$ is the phase difference produced by the current through the noise junction. Note that the Josephson
relation between the voltage and the phase from shot noise is different from that for the Josephson junction by the factor
of 2:
$V_s(t)=(\hbar/e) d\varphi_s(t)/dt_s$.  Then the probability of the forward tunneling is [instead of Eq. (\ref{ProbT})]
\begin{equation}
\Gamma^+(V) = {1\over \pi \hbar e^2 R_T}\mbox{Re}\left\{\int_0^\infty dt
\gamma(t)\left\langle\exp\left( J(t) + \frac{ieVt}{\hbar}+i\Delta \varphi_s\right) \right\rangle \right\} ~,
    \end{equation}
and the total current is 
\begin{eqnarray}
 I= { e}[\Gamma^+(V)-\Gamma^-(V)]= {1\over R_T} \left[
V + V_s  + {2\pi \over e\hbar \beta^2}\mbox{Im}\left\{  \int_{0}^\infty 
\frac{dt}{\sinh ^2 (\pi t/ \hbar \beta)} e^{J(t)}\left\langle\sin
\left( 
\frac{eVt}{\hbar}+\Delta \varphi_s\right) \right\rangle \right\}\right]~.
        \end{eqnarray}
Here $ V_s =I_sR$ is the average voltage generated by the current through the noise junction.  In
the
$T=0$ limit the current is:
\begin{eqnarray}
 I= {1\over R_T} \left[
V +V_s + {2\hbar  \over \pi e} \mbox{Im}\left\{ \int_{0}^\infty 
\frac{dt}{t^2} e^{J(t)}\left\langle\sin
\left( 
\frac{eVt}{\hbar}+\Delta \varphi_s\right) \right\rangle \right\}\right]~.
        \end{eqnarray}

For the normal junction one should replace $r$ by $2r$ in Eqs. (\ref{sin}) and (\ref{cos}) if the single-electron
quantum resistance $R_K$ is used in $r=\pi \rho=\pi R/R_K$. We need the approximate expressions for the phase
correlator at
$t\ll
\tau$:
\begin{eqnarray}
\langle \sin \Delta \varphi_s \rangle \approx {I_s \tau\over e}\left(\mbox{si} {2rt\over \tau }  +{\pi \over 2} 
\right)  ={I_s \tau\over e}\mbox{Si} {2rt\over \tau }  ~,            
\end{eqnarray}
\begin{eqnarray}
(\langle \cos \Delta \varphi_s \rangle-1)\approx {I_s \tau\over e}\left(\mbox{ci}
{2rt\over \tau } -\gamma-\ln {2rt\over \tau }\right)~.
              \end{eqnarray}
With help  of these expressions one receives the expressions for the ratchet  current,
\begin{eqnarray}
 I_0= {1\over R_T} \left[V_s + {2\hbar  \over \pi e}  \int_{0}^\infty 
\frac{dt}{t^2} e^{J_R(t)}\sin J_I(t)\left\langle \sin
\Delta \varphi(t_s,t)\right\rangle \right] \approx
 {1\over R_T} \left[V_s - {2\hbar  \over \pi e} {I_s \tau\over e}
\int_{0}^\infty 
\frac{dt}{t^2}  \sin {rt\over \tau }
\mbox{Si} {2 rt\over \tau }\right] \nonumber \\
={1\over R_T} \left[V_s - (1+\ln 2)I_s {\hbar  \over e^2} 
r\right]=0.153 {V_s\over  R_T}~,
        \end{eqnarray}
and for the shot-noise contribution to the conductance:
\begin{eqnarray}
G_s={d I\over dV}= {1\over R_T} \left[1 + {2 \over \pi }  \int_{0}^\infty 
\frac{dt}{t} e^{J_R(t)}\sin J_I(t)\left\langle\cos
\Delta \varphi(t_s,t)\right\rangle \right] \nonumber \\
\approx
-{1\over R_T}  {2 \over \pi } {|I_s| \tau\over e} \int_{0}^\infty 
\frac{dt}{t}\sin {rt\over \tau }\left(\mbox{ci}
{2rt\over \tau } -\gamma-\ln {2rt\over \tau }
\right)
={0.693\over
R_T}   {|I_s|\tau\over e} ={0.693\over
R_T}   {|V_s|\over e/C}~.
        \end{eqnarray}

\subsection{Response of the normal junction to an AC input}

The normal Coulomb blockade junction is sensitive to phase noise in an AC input as well. First let us consider the
linear response to a monochromatic input using analogy with the case of the Josephson junction (Sec. \ref{AClin}):
\begin{eqnarray}
  I= {1\over R_T} \left[V_i  + {2 \hbar \over \pi e }  \int_0^\infty 
\frac{dt}{t ^2 } \mbox{Im} \{e^{J_0(t)}\}\sin\Delta \varphi_i(t_0,t) \right]
\approx  {1\over R_T} \left[V_i  + {2 \hbar \over \pi e }  \int_0^\infty 
\frac{dt}{t ^2 } \mbox{Im} \{e^{J_0(t)}\}\Delta \varphi_i(t_0,t) \right]  \nonumber \\
={1\over R_T} \left[V_0 \cos \omega t_0  -\varphi_0 {2 \hbar \over \pi e }  \int_0^\infty 
\frac{dt}{t ^2 } \mbox{Im} \{e^{J_0(t)}\}[\cos \omega t_0\sin \omega t +\sin \omega t_0(1-\cos \omega
t)] \right]
= I_1 \cos \omega t_0 +I_2 \sin \omega t_0 ~,
     \end{eqnarray}
where $V_i=V_0 \cos \omega t_0$ with $V_0=\hbar \omega \varphi_0 /e$. The ohmic response is
\begin{eqnarray}
 I_1= {V_0\over R_T}  + {2\hbar  \over \pi e R_T}\varphi_0  \int_{0}^\infty 
\frac{dt}{t^2} \mbox{Im} \{e^{J_0(t)}\}\sin\omega t
= \varphi_0  I\left(\hbar \omega \over e\right)
=V_0{e\over \hbar \omega}I\left(\hbar \omega \over
e\right)~,
        \end{eqnarray}
where the $IV$ curve $I(V)$ of the normal junction without AC input was used [Eq. (\ref{i-v})] with $V$ replaced by $\hbar
\omega/e$.

There is also the reactive response:
\begin{eqnarray}
 I_2= {2\hbar  \over \pi e R_T}\varphi_0 \mbox{Im}\left\{ \int_{0}^\infty 
\frac{dt}{t^2} e^{J(t)}(1-\cos\omega t) \right\} \approx
{2\hbar  \over \pi e R_T}\varphi_0 \omega^2 \mbox{Im}\left\{\int_{0}^\infty 
dt  e^{J(t)}\right\} \nonumber \\
=-\omega V_0{2  \over \pi R_T}  \int_{0}^\infty 
dy  e^{-\pi \rho t/\tau}=-\omega V_0{2 \tau \over \pi^2 \rho R_T} 
=-\omega C V_0{2 R_K \over \pi^2 R_T} ~.
        \end{eqnarray}
The reactive response is a result of the Coulomb blockade correction to the geometric capacitance:
\begin{eqnarray}
 \Delta C= C {2 R_K \over \pi^2 R_T}
        \end{eqnarray}
similar to the correction for the Josephson junction, Eq. (\ref{Ccor}).

Quadratic correction to the DC zero-bias conductance is also sensitive to the decoherence time $\tau_c$:
\begin{eqnarray}
\Delta G\approx -{1\over R_T} {1 \over \pi } \mbox{Im}\left\{ \int_{0}^\infty 
\frac{dt}{t} e^{J_0(t)}\left\langle\Delta \varphi_i^2\right\rangle \right\}
=-{1\over R_T} {\varphi_0^2 \over \pi } \mbox{Im}\left\{ \int_{0}^\infty 
\frac{dt}{t} e^{J_0(t)}(1-\cos \omega t e^{-t/\tau_c})\right\} \nonumber \\
\approx -{1\over R_T} {\varphi_0^2 \over \pi \tau_c} \mbox{Im}\left\{ \int_{0}^\infty 
dt e^{J_0(t)}\right\} = {1\over R_T} {\varphi_0^2 \over \pi \tau_c} \int_{0}^\infty 
dy e^{-\pi \rho/\tau}= {1\over R_T} {\varphi_0^2 \tau\over \pi^2 \rho \tau_c} 
 = {2\over R_T} {V_0^2 C\over \pi\hbar \omega^2 \tau_c}~.
        \end{eqnarray}

\section{Josephson junction, strong coupling limit, shot noise} \label{sec5}

\subsection{Without shot noise}

\begin{figure}
  \begin{center}
    \leavevmode
    \includegraphics[width=0.8\linewidth]{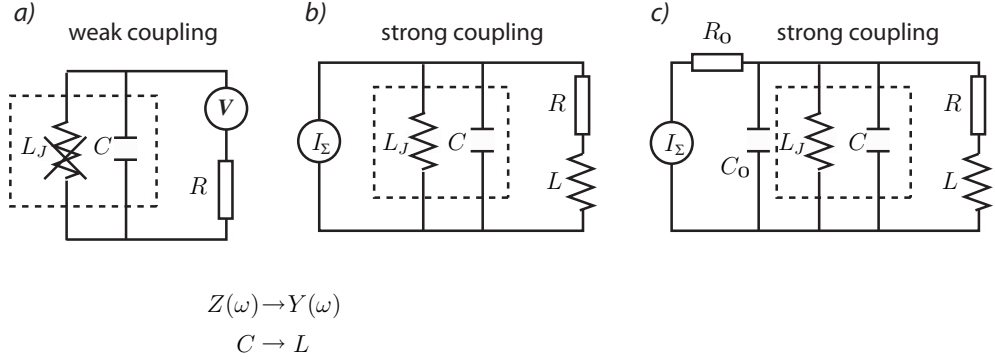}
    \bigskip
    \caption{Strong coupling limit and environment duality. The dashed squares 
show the circuit elements, which present the Josephson and the noise junctions with  the capacitance $C=C_J+C_s$. The
current source $I_\Sigma=I+I_s$ provides the constant current $I$ and the noisy current $I_s$ in parallel. a)
The dual weak-coupling case.  The inductance
$L_J$ of the Josephson junction is negligible in the weak-coupling limit. b) The dual strong-coupling limit. c) The
strong-coupling case taking into account the capacitance
$C_0$ and the resistance $R_0$ of the leads.}
  \label{fig3}
  \end{center}
  \end{figure}

 In the case of equilibrium noise the strong coupling limit can be investigated using the
duality relations, which connect the parameters of the Josephson junction in the weak coupling and the strong
coupling limits \cite{SZ,Weiss}:
\begin{equation}
E_J \rightarrow 2 \Delta~,~~V \rightarrow {h \over 4e^2}I = R_Q I~,~~\varphi
\rightarrow {\pi Q \over e}~,~~~\rho={R\over R_Q} \rightarrow \tilde
\rho={R_Q\over R}~,~~~Z \rightarrow Y~.
      \label{du} \end{equation}
Here $Y(\omega)$ is the admittance of the electric circuit, and $\Delta$ is the matrix element for a transition between adjacent
states of the Josephson junction with localized phases $2\pi n$ and $2\pi (n+1)$ ($\hbar/\Delta$ is the life time at the state,
which corresponds to some localized phase). Using these relations the
$IV$ curve in the weak coupling limit, Eq. (\ref{WC}), transforms to the dual $VI$ curve in the strong-coupling
limit:
\begin{eqnarray}
V=-{4\pi \Delta^2 \over e \hbar}\int_0^\infty dt \mbox{Im} \left\{e^{\tilde J_0(t)}
\right\} \sin \left(2eR_Q It\over \hbar\right)
=-{4\pi \Delta^2 \over e \hbar}\int_0^\infty dt \mbox{Im} \left\{e^{\tilde J_0(t)}
\right\} \sin \left(\pi It\over e\right)~.
 \end{eqnarray}
The charge-charge correlator [see Eq. (149) in Ref. \onlinecite{IN}] is given by the relation dual to Eq.
(\ref{phaseCor}):
\begin{eqnarray}
\tilde J_0(t)=2R_Q  \int_0^\infty
{d\omega \over \omega} {\mbox{Re}  Y(\omega)}
 \left[\coth\left({1\over 2}\beta \hbar \omega\right)(\cos \omega t -1)
-i\sin \omega t\right] ~.
    \label{Q-IN}  \end{eqnarray}
 
In order to exploit duality completely, i.e., not only for the general expression for the $IV$ curve but also for the
specific results for ohmic environment with the effective impedance $Z=(1/R+i\omega C )^{-1}$, one should find a
corresponding ``dual'' environment for the strong-coupling case. In
the weak-coupling case the finite relaxation time $\tau=RC$ is provided by
the imaginary element of the impedance $Z$, which is the capacitance $C=C_J+C_s$ of the two junctions (without the
noise source $C_s=0$). In the strong coupling case the both imaginary elements of the Josephson junction (capacitance
and inductance) do not contribute to the admittance (see Fig. \ref{fig3}b). In order to get a finite relaxation time
dual to the time $\tau=RC$ of the weak coupling limit one should add to the effective electric circuit
the inductance $L$ in series with the ohmic shunt resistance $R$ (Fig. \ref{fig3}b). Then the admittance
dual to the impedance $Z$ of the weak coupling limit is 
\begin{equation}
 Y(\omega)=\frac{1}{R +i\omega L}=\frac{1/R}{1+i\omega \tau_L}~,
      \end{equation}
where the circuit relaxation time $\tau_L =L/R$ is dual to the relaxation time $\tau=RC$ in the weak
coupling limit. Thus duality relations Eq. (\ref{du}) should by supplemented by the ``environment
duality'' relation $C \rightarrow L$. Note, however, that introduction of finite inductance $L$ is not
the only way to achieve a finite relaxation time. For example, Sch\"on and Zaikin \cite{SZ} suggested
that the period of Josephson plasma oscillation $\sim \hbar/\sqrt{E_J E_c}$ should play the role of
relaxation time dual to the time $RC$ in the weak coupling limit. 

\subsection{With shot noise}

The duality concept is valid only for equilibrium noise when symmetry
between current and voltage or impedance and admittance takes place. In the case of shot noise this
symmetry is broken since shot noise is a {\em current} noise, which produces asymmetric effects on the
current and on the voltage. Though we can still use the duality for description of the equilibrium
contribution to the noise (this contribution cannot be ignored even if our goal is the effect of shot
noise), the effect of shot noise on the $IV$ curve is essentially different in the weak and the strong
coupling limits.

In the strong-coupling limit the effect of environment is governed by charge fluctuations. Whereas the phase
fluctuations were determined by the voltage fluctuations via the Josephson relation $\hbar d\varphi/dt=2eV$, the charge
fluctuations are determined by the current fluctuations via the relation
$dQ/dt =I$. Important is that the current $I$ should be the {\em total current} of the part of the circuit, which describes the
Josephson junction in the RCSJ model (inductance + capacitance + shunt resistance). Single-electron tunneling through the noise
junction produces a pulse  of the total current. If the tunneling time is very short, the pulse can be approximated by the
$\delta$-function current peak. The effective circuit in  Fig. \ref{fig3}b cannot broaden the original $\delta$-function
current peak. This is in contrast to the voltage pulses, which were relevant for phase fluctuations in the weak coupling
limit: the $\delta$-function current peak produced the voltage pulse of width $\tau=RC$.
Meanwhile, the finite width of the current pulse is of principal importance for the shot-noise effect  and therefore cannot be
ignored. Broadening of the current pulse can be determined by additional lump elements of the electric circuit, which are
connected with the ohmic resistance $R_0$ and the capacitance $C_0$ of wires connecting the current-noise source and the probe
Josephson junction (see Fig. \ref{fig3}c). Bearing in mind this electric circuit the single-electron tunneling at $t-0$ produces
the current pulse
\begin{equation}
\delta I (t)=\mbox{sign}(I_s){e\over \tau_s} e^{-t/\tau_s}~,
      \end{equation}
which brings the charge
\begin{equation}
\delta Q_s(t)=\mbox{sign}(I_s)e(1- e^{-t/\tau_s})~.
      \end{equation}
Here $\tau_s=R_0C_0$. 
The $VI$ curve with shot noise is given by
\begin{eqnarray}
V=-{4\pi \Delta^2 \over e \hbar}\int_0^\infty dt \mbox{Im} \left\{e^{\tilde J_0(t)}
\right\} \left\langle\sin \left({\pi It\over e}+{\pi\Delta Q_s\over e}\right)\right\rangle~,
 \end{eqnarray}
where $\Delta Q_s=Q_s(t_0)-Q_s(t_0-t)$ and
\begin{eqnarray}
Q_s(t)  =\mbox{sign}(I_s)\pi \sum_i \Theta (t-t_i)\left [ 1-
e^{-(t-t_i)/\tau}\right]  ~.
     \end{eqnarray}
The shot-noise charge-charge correlators are found by averaging over the the time $t_0$. 
 The expressions for $\langle\cos (\pi \Delta Q_s/e)\rangle$ and $\langle\sin
(\pi \Delta Q_s/e)\rangle$ follow from those for $\langle\cos (\Delta
\varphi_s)\rangle$ and $\langle\sin (\Delta \varphi_s)\rangle$ assuming $\rho=1$ and using the duality
relations between the phase and the charge:
\begin{eqnarray}
{e\over I_s \tau_s}\left\langle \sin {\pi\Delta Q_s\over e}\right\rangle = 
\mbox{si} \pi - \mbox{si} \left(\pi
e^{-t/\tau_s} \right) +  \mbox{si} \left[\pi\left(1-
e^{-t/\tau_s} 
\right) \right] +{\pi \over 2}~,
       \label{sinG}       \end{eqnarray}
\begin{eqnarray}
{e\over |I_s |\tau_s}\left(\left\langle \cos {\pi\Delta Q_s\over e} \right\rangle-1\right)= -
\mbox{ci} \pi + \mbox{ci} \left(\pi  e^{-t/\tau_s} \right)-{t\over
\tau_s} +
\mbox{ci} \left[\pi\left(1- e^{-t/\tau_s} 
\right) \right] -\gamma-\ln \left[\pi\left(1-
e^{-t/\tau_s} 
\right) \right] ~.
    \label{cosS}           \end{eqnarray}
In contrast to the weak-coupling case where the single relaxation time $\tau$ characterized both the
equilibrium and the shot-noise correlators, in the strong coupling case the equilibrium charge correlator is
characterized by the time $\tau_L$, whereas the shot-noise charge correlator has the other characteristic time
$\tau_s$. We expect that $\tau_s \ll \tau_L$. Therefore we should consider the limit $t \gg \tau_s$ in Eqs.
(\ref{sinG}) and (\ref{cosS}):
\begin{eqnarray}
\left\langle \sin {\pi\Delta Q_s\over e}\right\rangle =(2 \mbox{si} \pi  +  \pi) {I_s\over e} \tau_s
 =3.704 {I_s\over e} \tau_s~,
              \end{eqnarray}
\begin{eqnarray}
\left\langle \cos {\pi\Delta Q_s\over e} \right\rangle-1= -2{|I_s|\over e} t ~.
         \end{eqnarray}

We expand the shot noise contribution with respect to the current $I$:
\begin{eqnarray}
\Delta V=-{4\pi \Delta^2 \over e \hbar}\int_0^\infty dt \mbox{Im} \left\{e^{\tilde
J_0(t)}
\right\}\left[ \sin \left({\pi It\over e}\right) \left(\left\langle\cos{\pi\Delta Q_s\over e}
\right\rangle-1\right) +\cos \left({\pi It\over e}\right) \left\langle\sin{\pi\Delta Q_s\over e}
\right\rangle\right]\nonumber \\
\approx -{4\pi \Delta^2 \over e \hbar}\int_0^\infty dt \mbox{Im} \left\{e^{\tilde
J_0(t)}
\right\}\left[ {\pi It\over e} \left(\left\langle\cos{\pi\Delta Q_s\over e}
\right\rangle-1\right) +\left\langle\sin{\pi\Delta Q_s\over e}
\right\rangle\right] = V_0+R_s I ~.
 \end{eqnarray}
Rotating the integration path in the complex plane ($t\to -iy$)  we obtain the
ratchet voltage $V_0$ and the zero-bias resistance $R_s $:
\begin{eqnarray}
V_0 =-{4\pi \Delta^2 \over e \hbar}\int_0^\infty dt \mbox{Im} \left\{e^{\tilde
J_0(t)}
\right\}\left\langle\sin{\pi\Delta Q_s\over e}
\right\rangle
\nonumber \\
=-{14.8\pi \Delta^2 \over e^2 \hbar}I_s \tau_s\int_0^\infty dt \mbox{Im}
\left\{e^{\tilde J_0(t)}
\right\}={14.8\pi \Delta^2 \over e^2 \hbar}I_s \tau_s\int_0^\infty dy
e^{-\pi\tilde \rho y/\tau_L}={14.8 \Delta^2 \over e^2
\hbar}{ \tau_L\over \tilde \rho }I_s \tau_s={7.4 \pi\Delta^2 \over e^4}{L\tau_s\over R_Q^2 }I_s ~,
 \end{eqnarray}
\begin{eqnarray}
R_s =-{4\pi^2 \Delta^2 \over e^2 \hbar}\int_0^\infty t\, dt \mbox{Im} \left\{e^{\tilde
J_0(t)}
\right\}\left(\left\langle\cos{\pi\Delta Q_s\over e}
\right\rangle-1\right) 
\nonumber \\
=| I_s| {8\pi^2 \Delta^2 \over e^3 \hbar}\int_0^\infty t^2\,dt \mbox{Im}
\left\{e^{\tilde J_0(t)}\right\}=| I_s| {8\pi^2 \Delta^2 \over e^3 \hbar}\int_0^\infty
y^2\,dy e^{-\pi\tilde \rho y/\tau_L} =| I_s| {16\pi^2 \Delta^2 \over e^3 \hbar}\left(\tau_L \over \pi
\tilde
\rho\right)^3 =| I_s| {8 \Delta^2 \over  e^5 }{L^3 \over R_Q^4}~. 
 \end{eqnarray}
One can see that the even effect (resistance $R_s$) remains finite in the limit $\tau_s \to 0$. In contrast, the
ratchet effect, which is an odd effect, vanishes. More generally, any odd effect should vanish in the limit $\tau_s
\to 0$. This is because in this limit tunneling events lead to instantaneous transfers of charge $\pm e$. The odd
effect should depend on the sign of transferred charge, but the current cannot depend on whether the charge $+e$ or
$-e$ suddenly appears at the Josephson junction due to periodicity of the charge correlator with the period
$2e$: the instantaneous transfers of the charges $+e$ and $-e$ must produce identical effects. Only finite duration of
the current pulse (finite $\tau_s$) can lead to an odd effect (ratchet voltage $V_0$ as an example).

This interesting feature of the strong-coupling case becomes even more pronounced if shot noise originates
from the current of Cooper pairs. This can be realized if the noise junction is a SIN junction in the regime of
the Andreev reflection or a Josephson junction in the weak coupling limit in the regime of incoherent tunneling
of Cooper pairs. In the both cases the tunneling event is accompanied by the charge jump
$Q_s=2e(1-e^{-t/\tau_s})\mbox{sign}(I_s)$ and, in analogy with Eqs. (\ref{sinG}) and (\ref{cosS}), the odd and
the even charge correlators can be obtained from the phase-phase correlators using duality and assuming
$\rho=2$:
\begin{eqnarray}
{e\over I_s \tau_s}\left\langle \sin {\pi\Delta Q_s\over e}\right\rangle = 
-\mbox{si}(2 \pi) + \mbox{si} \left(2\pi e^{-t/\tau_s} \right) +  \mbox{si} \left[2\pi\left(1-
e^{-t/\tau_s} \right) \right] +{\pi \over 2}~,
              \end{eqnarray}
\begin{eqnarray}
{e\over |I_s| \tau_s}\left(\left\langle \cos {\pi\Delta Q_s\over e} \right\rangle-1\right)= \mbox{ci}(2 \pi) -
\mbox{ci} \left(2\pi  e^{-t/\tau_s} \right)-{t\over
\tau_s} +\mbox{ci} \left[2\pi\left(1- e^{-t/\tau_s} \right) \right] -\gamma-\ln \left[2\pi\left(1-
e^{-t/\tau_s} \right) \right] ~.
               \end{eqnarray}
In the limit $t\gg \tau_s$
\begin{eqnarray}
{e\over I_s \tau_s}\left\langle \sin {\pi\Delta Q_s\over e}\right\rangle =2\pi  e^{-t/\tau_s}~,
              \end{eqnarray}
\begin{eqnarray}
{e\over |I_s |\tau_s}\left(\left\langle \cos {\pi\Delta Q_s\over e}\right \rangle-1\right) 
= 2[\mbox{ci} (2\pi) - \gamma-\ln
(2\pi)] +\pi^2e^{-2t/\tau_s} ~.
               \end{eqnarray}
Calculating the shot-noise contribution to
the voltage one cannot expand in $t/\tau_s$ but can expand in
$t/\tau_L$ and in the current $I$:
\begin{eqnarray}
\Delta V =-{4\pi \Delta^2 \over e \hbar}\int_0^\infty dt \mbox{Im} \left\{e^{\tilde J_0(t)}
\right\}\left[ \left(\left \langle \cos{\pi\Delta Q_s\over e}\right\rangle-1\right))\sin \left({\pi It\over e}\right) +\left
\langle
\sin {\pi\Delta Q_s\over e}\right \rangle\cos \left({\pi It\over e}\right) \right] \nonumber \\
\approx {4\pi^2 \Delta^2 \over e \hbar}{\tilde \rho\over \tau_L}\int_0^\infty t\,dt \left[\left (\left \langle
\cos
{\pi\Delta Q_s\over e}\right \rangle-1\right){\pi It\over e} +\left \langle \sin {\pi\Delta Q_s\over e}
\right\rangle\right]=V_0+R_s I~.
 \end{eqnarray}
Integrating by parts one obtains the ratchet voltage 
\begin{eqnarray}
V_0 = {4\pi^2 \Delta^2 \over e \hbar}{\tilde \rho\over \tau_L}\int_0^\infty t\,dt \left\langle \sin
{\pi\Delta Q_s\over e} \right \rangle={8\pi^3 \Delta^2 \over e^2 \hbar}{\tilde \rho\over \tau_L}I_s
\int_0^\infty  e^{-t/\tau_s}{t^2\over 2}\,dt \left\{\mbox{si}'(2\pi e^{-t/\tau_s}) \right. \nonumber \\ \left.
-\mbox{si}'\left[2\pi\left(1- e^{-t/\tau_s} \right) \right]\right\} 
={2\pi^2 \Delta^2 \over e^2 \hbar}{\tilde \rho\over \tau_L}I_s
\tau_s^3\int_0^\infty x^2\,dx \left[\sin(2\pi e^{-x})-{\sin\left[2\pi\left(1- e^{-x} \right) \right]\over
e^x-1}\right]={7.62\pi^3 \Delta^2  \tau_s^3\over e^4  L}  I_s ~,
 \end{eqnarray}
and the short noise contribution to the zero-bias resistance
\begin{eqnarray}
R_s = {4\pi^3 \Delta^2 \over e^2 \hbar}{\tilde \rho\over \tau_L}\int_0^\infty t^2\,dt  \left(\left\langle
\cos{\pi\Delta Q_s\over e} \right\rangle-1\right)=- {4\pi^3 \Delta^2 \over e^3 \hbar}{\tilde \rho\over
\tau_L}|I_s|\int_0^\infty {t^3\over 3}\,dt \left[ \mbox{ci}'(2\pi e^{-t/\tau_s})2\pi e^{-t/\tau_s}-1
\right.\nonumber \\ \left. +\mbox{ci}'\left[2\pi\left(1- e^{-t/\tau_s} \right) \right]2\pi
e^{-t/\tau_s} -{e^{-t/\tau_s} \over 1-e^{-t/\tau_s}}\right]\nonumber \\
= {4\pi^3 \Delta^2 \tau_s^4\over 3e^3 \hbar}{\tilde \rho\over \tau_L}|I_s|\int_0^\infty x^3\,dx
\left[1-\cos(2\pi e^{-x})  +{1-\cos\left[2\pi\left(1- e^{-x} \right) \right]\over e^x-1}\right] 
 ={5.08\pi^4\Delta^2  \tau_s^4\over
e^5L}|I_s| ~.
 \end{eqnarray}
Thus shot noise from the current of Cooper pairs produces neither odd nor even effects
if duration $\tau_s$ of a single current pulse vanishes.

What would happen if experimentalists were able to decrease the time $\tau_s$ essentially? Eventually this
time would be comparable or even less that the intrinsic time, which determines duration of
single-electron tunneling. Then the experimental observation of the zero-bias anomaly would provide an
information on this intrinsic time. Up to now the tunneling time is mostly an issue of theoretical debates
\cite{trav,Hauge,Land}: Though some attempts to observe it were undertaken, but their interpretation is ambiguous (see
Ref. \onlinecite{Land} and references therein). The subject is controversial and the consensus on how this time should
be defined is absent. Decreasing of $R_0$ and
$C_0$ the experimentalists would shift this issue  from the pure theory to the measurement. How realistic could this
perspective be? Presently it is difficult to achieve
$R_0$ less than 100 $\Omega$ and $C_0$ less then 1 fF. This yields $\tau_s=R_0C_0 >10^{-13}$ s. On the other hand,
whatever disagreement on definition of the tunneling time, usually it is estimated as
$10^{-15}~-~10^{-14}$ s. This yields the hope that further progress in miniaturization of the
experimental set up or searching for cases with longer tunneling time could open a new possibility to investigate
the tunneling time experimentally.

\section{Conclusions}

We have presented the theory of the  effect of shot noise from an independent source on the Coulomb blockaded
Josephson junction in high-impedance environment. The analysis takes into account asymmetry of 
shot noise characterized by its odd moments. For high impedance environment the effect is so strong that the expansion in
moments is not valid and was not used in the analysis. Asymmetry of shot noise results in asymmetry of the $IV$
curve: the shift of the conductance minimum from the zero bias and the ratchet effect, which
have been observed experimentally \cite{SN,exp}. At low noise currents (currents responsible for shot noise) the effect is
proportional to the noise current independently of counting statistics of electron transport. However, at high noise
currents the effect of noise on the $IV$ curve is very sensitive to electron counting statistics. This high
sensitivity is explained by the fact that in contrast to usual methods of noise detection the Coulomb
blockaded junction probes phase but not voltage.

The theory was generalized on  another type of noise (phase noise of a monochromatic AC input) and on a normal
Coulomb blockaded tunnel junction, In the both cases the effect of noise strongly affects the Coulomb
zero-bias anomaly of the $IV$ curve. The effect of shot noise on the superconducting Josephson junction in
low-impedance environment was also considered. If environment generates only the equilibrium Johnson-Nyquist
noise this case can be analyzed using the duality  relations with the case of the Coulomb blockaded Josephson
junction in high-impedance environment. However, shot noise breaks duality relations between the two cases. An
interesting feature of the superconducting Josephson junction in low-impedance environment has been revealed:
the effect of shot noise on its $IV$ curve can give information on the time of electron tunneling through the
junction responsible for shot noise.

Altogether, the analysis demonstrates, that the  low-bias part of the $IV$ curves of tunnel junctions, both Coulomb
blockaded and superconducting, can be exploited as a
sensitive detector of various types of noise.

\section*{Acknowledgements}

The author appreciates collaboration and discussions with Julien Delahaye, Pertti Hakonen,
Tero Heikkil\"a, Rene Lindell, Mikko Paalanen, Jukka Pekola, and Mikko Sillanp\"a\"a. The author also thanks Markus
B\"uttiker, Rosario Fazio, and Bertrand Reulet for interesting discussions of the results of the paper. The work was supported by 
the Large Scale Installation Program ULTI-3 of the European Union and by the grant of the Israel Academy of Sciences and
Humanities.

\end{document}